\documentclass[12pt]{article}
\usepackage{a4wide}

\usepackage{amsmath}
\usepackage{amsfonts}
\usepackage{amscd}
\usepackage{amsthm}
\usepackage{mathtools}

\usepackage{slashed}
\usepackage{graphicx}

\usepackage{cite}

\usepackage[nolist,printonlyused]{acronym}

\usepackage{pslatex}
\usepackage[latin1,utf8]{inputenc}
\usepackage[english]{babel}
\usepackage[OT2,T1]{fontenc}

\usepackage[babel]{csquotes}

\allowdisplaybreaks

\newcommand{\diff}{\mathrm{d}}     
\newcommand{\alphaS}{\alpha_\mathrm{s}} 
\newcommand{\MSbar}{\overline{\text{MS}}} 
\newcommand{\CF}{C_\mathrm{F}^{}}

\newcommand{\bq}{\begin{eqnarray}}
\newcommand{\eq}{\end{eqnarray}}

\usepackage{color}

\begin{document}

\begin{acronym}

\acro{sm}[SM]{Standard Model}

\acro{qft}[QFT]{quantum field theory}

\acro{qed}[QED]{quantum electrodynamics}

\acro{qcd}[QCD]{quantum chromodynamics}

\acro{rge}[RGE]{renormalization group equation}

\acro{lo}[LO]{leading order}

\acro{nlo}[NLO]{next-to-leading order}

\acro{nnlo}[NNLO]{next-to-next-to-leading order}

\acro{uv}[UV]{ultraviolet}

\acro{ir}[IR]{infrared}

\acro{ms}[MS]{minimal subtraction}

\acro{msbar}[$\MSbar$]{modified minimal subtraction}

\acro{1pi}[1PI]{One particle irreducible}

\acro{lhc}[LHC]{Large Hadron Collider}

\acro{kln}[KLN]{Kinoshita--Lee--Nauenberg}

\acro{mc}[MC]{Monte Carlo}

\acro{scet}[SCET]{soft-collinear effective theory}

\acro{hqet}[HQET]{heavy quark effective theory}

\acro{bhqet}[bHQET]{boosted heavy quark effective theory}

\acro{nll}[NLL]{next-to-leading logarithmic}

\acro{nnll}[NNLL]{next-to-next-to-leading logarithmic}

\end{acronym}

\thispagestyle{empty}

\begin{flushright}
  MITP/20-021
\end{flushright}

\vspace{1.5cm}

\begin{center}
  {\Large\bf Cutoff dependence of the thrust peak position in the dipole shower\\
  }
  \vspace{1cm}
  {\large Robin Baumeister and Stefan Weinzierl\\
\vspace{2mm}
      {\small \em PRISMA Cluster of Excellence, Institut f{\"u}r Physik, }\\
      {\small \em Johannes Gutenberg-Universit{\"a}t Mainz,}\\
      {\small \em D - 55099 Mainz, Germany}\\
  } 
\end{center}

\vspace{2cm}

\begin{abstract}\noindent
  {
We analyse the dependence of the peak position of the thrust distribution on the cutoff value in the Nagy-Soper dipole shower. 
We compare the outcome of the parton shower simulations to a relation of the dependence from an analytic computation,
derived within soft-collinear effective theory.
We show that the result of the parton shower simulations and the analytic computation are in good agreement.
   }
\end{abstract}

\vspace*{\fill}

\newpage
\section{Introduction}
\label{sect:intro}

Parton showers are widely used in event generators like Pythia \cite{Sjostrand:2004ef,Sjostrand:2014zea}, Herwig \cite{Gieseke:2003rz,Bellm:2015jjp} or Sherpa \cite{Krauss:2001iv,Krauss:2005re,Gleisberg:2008ta}.
There are several variants of parton showers available (angular ordered showers \cite{Marchesini:1983bm,Webber:1983if}, 
dipole showers \cite{Gustafson:1986db,Gustafson:1987rq,Andersson:1989ki,Andersson:1988gp,Lonnblad:1992tz,Lonnblad:2001iq,Nagy:2005aa,Nagy:2006kb,Dinsdale:2007mf,Schumann:2007mg,Platzer:2011bc,Nagy:2014mqa,Hoche:2015sya,Dasgupta:2018nvj,Forshaw:2020wrq,Dasgupta:2020fwr}, antenna showers \cite{Giele:2007di,Fischer:2016vfv,Fischer:2017htu} etc.).
Common to all parton showers is the dependence on two important parameters:
The coupling, usually specified by the value $\alphaS(m_Z)$ and the cutoff scale $Q_0$.
While the physical meaning of the former parameter is immediately clear -- it describes the strength of the interaction --,
the dependence on the latter parameter has received recently more attention \cite{Hoang:2018zrp}.
This has been triggered to some extent by the desire to understand top quark mass measurements \cite{Hoang:2008xm,Hoang:2014oea,Nason:2017cxd,Corcella:2017rpt,Heinrich:2017bqp,Ravasio:2018lzi}.
In the extraction of the top mass from experimental measurements theory sneaks in through the use of the template method or the matrix element method.
For example, within the template method one generates first by Monte Carlo a sample of events for various values of $m_{\mathrm{MC}}$
and then determines the best fit to the experimental data.
The Monte Carlo mass $m_{\mathrm{MC}}$ is only implicitly defined through the program code of the event generator.
We would like to relate the Monte Carlo mass $m_{\mathrm{MC}}$ to a theoretically well-defined mass.
In order to avoid renormalon ambiguities the theory mass should be a short-distance mass.
The $\mathrm{MSR}$-mass \cite{Hoang:2008yj,Hoang:2017suc} is a good candidate.
The $\mathrm{MSR}$-mass depends on two scales,
a UV-scale $\mu$ and an IR-scale $R$, such that
\bq
 m_{\mathrm{MSR}}\left(R=0\right)
 =
 m_{\mathrm{pole}},
 & &
 m_{\mathrm{MSR}}\left(R=\bar{m}\right)
 =
 \bar{m}.
\eq
Here, $\bar{m}$ denotes the $\overline{\mathrm{MS}}$-mass. The $\mathrm{MSR}$-mass is a short-distance mass for $R \gg \Lambda_{\mathrm{QCD}}$.
The definition of the $\mathrm{MSR}$-mass can be understood in the context of effective theories, which separates the various relevant scales
(the centre-of-mass energy $Q$, the top mass $m_t$, the top width $\Gamma_t$ and $\Lambda_{\mathrm{QCD}}$).
It can be shown that physics at scales in the range of $[\Gamma_t,m_t]$ affect
the position of the peak of the reconstructed top mass distribution, and these effects depend on $m_t$.
Physics at scales in the range of $[\Lambda_{\mathrm{QCD}},\Gamma_t]$ affect the peak position as well, but these effects are described by soft physics and 
are independent of $m_t$.
Thus $R \approx \Gamma_t$ is a natural choice for the infrared scale.
It is a numerical coincidence that the shower cutoff is typically of the order of $\Gamma_t$.
Furthermore the shower cutoff provides naturally an infrared scale for the parton shower.
This implies that the Monte Carlo mass $m_{\mathrm{MC}}$ is closely related to the $\mathrm{MSR}$-mass with the IR-scale $R$ being of the order of
the shower cutoff $Q_0$.

In an idealised perfect event simulator an observable like the peak position of the reconstructed top mass distribution should be independent
of a technical parameter like the shower cutoff.
The explicit dependence on the shower cutoff will be compensated by a change in the modelling of soft physics
and by a change of the IR-scale $R$ in the top mass definition.
This dependence is in principle calculable analytically, and keeping the modelling of soft physics and the IR-scale $R$ fixed one obtains a prediction
how the peak position of the reconstructed top mass distribution in simulated events depends on the shower cutoff.
 
In order to understand the situation better, it is helpful to consider a related problem: 
The dependence of the peak position of the thrust distribution in electron-positron annihilation on the shower cutoff \cite{Fleming:2007xt}.
This quantity has the advantage that analytic results are available for the dependence on the shower cutoff (contrary to the case discussed above).
The analytic results may be derived within soft-collinear effective theory or alternatively within the coherent branching formalism \cite{Marchesini:1983bm,Marchesini:1987cf,Catani:1990rr,Gieseke:2003rz}.
In a recent publication \cite{Hoang:2018zrp} a study has been performed which compares the analytic results with numerical results
from an angular ordered shower as implemented in Herwig.
Clearly, the coherent branching formalism is closely tied to an angular ordered shower.
However, the analytic results in \cite{Hoang:2018zrp} have been derived with the assumption that the shower cutoff provides a restriction
on the transverse momentum of the emitted particles. 
Furthermore, the second derivation in \cite{Hoang:2018zrp} based on soft-collinear effective theory uses just the condition on the transverse momentum
and is not tied to any particular shower algorithm.
We thus expect these results to hold for any shower algorithm whose shower evolution variable reduces in the singular limit to $p_\perp$.
This is a hypothesis which can be tested and verified by a dedicated study: This is the content of the present paper.
We show that the theoretical prediction is not only valid for angular ordered parton showers 
but holds for Nagy-Soper dipole showers (with $p_\perp$ as evolution variable) as well.
By a Nagy-Soper dipole shower we mean a shower algorithm as proposed in refs.~\cite{Nagy:2005aa,Nagy:2006kb}.
As an analytic treatment of the Nagy-Soper dipole shower is out of reach, we do this by comparing the theoretical prediction
with numerical results from shower simulations.
Comparing the results of \cite{Hoang:2018zrp} with our results we observe that the dipole shower agrees slightly better
with the theoretical predictions.

This paper is organised as follows: In the next section we review the analytic computations for the cutoff dependence and give the essential relation. In section \ref{sect:dipoleShower} we present our analysis and results using the dipole shower. Finally, our conclusions are contained in section \ref{sect:conclusions}.

\section{Analytic results}
\label{sect:analytic}

We study the observable ``one minus thrust'' $\tau$ in electron-positron annihilation, defined by
\bq
 \tau & = & 1 - T.
\eq
The thrust is defined by
\bq
\label{def_thrust}
 T & = & 
 \frac{1}{Q} \max\limits_{\vec{n}} \sum\limits_j \left| \vec{p}_j \cdot \vec{n} \right|,
\eq
where $Q$ denotes the centre-of-mass energy,
$\vec{p}_j$ denotes the three-momentum of particle $j$ and the sum runs over all particles in the 
final state. The thrust variable maximises the total longitudinal momentum along the unit vector $\vec{n}$.
We will study this observable in massless QCD, where the leading order Born process is
$e^+ e^- \rightarrow Z/\gamma \rightarrow \bar{q} q$, with massless quarks $q$.
We also study this observable in top-pair production in electron-positron annihilation,
where the leading order Born process is
$e^+ e^- \rightarrow Z/\gamma \rightarrow \bar{t} t$.
In the former case (massless QCD) the definition~(\ref{def_thrust}) agrees
with the conventional definition of thrust \cite{Brandt:1964sa,Farhi:1977sg}
\bq
 T & = & 
 \frac{\max\limits_{\vec{n}} \sum\limits_j \left| \vec{p}_j \cdot \vec{n} \right|}{\sum\limits_j \left| \vec{p}_j \right|},
\eq
in the latter case (massive quark production) with the definition used in \cite{Fleming:2007qr,Stewart:2010tn}.
For the analysis, the peak position of the $\tau$ distribution, $\tau_\mathrm{peak}$, is considered. 
It is strongly affected by configurations containing two jets that are back-to-back. 
These contributions arise from the leading order production of a quark-antiquark pair.
For massless quark production, the region of the peak is close to $\tau=0$. 
This location is shifted to positive values due to non-perturbative effects. 
The size of this shift is of order $\mathcal{O}(\Lambda/Q)$, where $\Lambda\approx 1$ GeV \cite{Dokshitzer:1995zt,Dokshitzer:1995qm,Dokshitzer:1997ew,Abbate:2010xh}.
For massive quark production, the peak is located around $\tau = 2 m^2/Q^2$ \cite{Hoang:2018zrp}.
The thrust distribution makes a reasonable choice for the study of the cutoff dependence of a parton shower 
because there exist analytic calculations for the thrust distribution based on factorisation, 
that the parton shower outcome can be compared to.

The hadronic thrust distribution in the peak region is given by \cite{Korchemsky:1999kt,Korchemsky:2000kp,Abbate:2010xh}
\bq
\label{eq:HadronicThrust}
 \frac{\diff\sigma}{\diff\tau}(\tau,Q) 
 & = & 
 \int\limits_0^{\tau Q} \diff\mu\, \frac{\diff\sigma_{\mathrm{sing}}}{\diff\tau}\left(\tau-\frac{\mu}{Q},Q\right)\, S_\mathrm{mod}(\mu),
\eq
where $\diff\sigma_{\mathrm{sing}}/\diff\tau$ contains the resummed 
singular partonic QCD corrections (i.e. terms corresponding to $\alphaS^n \delta(\tau)$ and $\alphaS^n[\ln^k(\tau)/\tau]_+$)
and the soft model shape function $S_\mathrm{mod}$ describes soft-gluon non-perturbative effects. 
In our numerical studies we substitute $\diff\sigma_{\mathrm{sing}}/\diff\tau(\tau,Q)$
by $\diff\sigma_{\mathrm{shower}}/\diff\tau(\tau,Q,Q_0)$.
The latter is obtained from a simulation with hard matrix elements at the scale $Q$ and shower evolution from the hard scale $Q$ to
the shower cutoff scale $Q_0$.

\subsection{The massless case}

We start with the case of massless QCD.
The leading order Born process is
$e^+ e^- \rightarrow Z/\gamma \rightarrow \bar{q} q$, with massless quarks $q$.
We take the same soft model shape function $S_\mathrm{mod}$ as in \cite{Hoang:2018zrp}
\bq
\label{eq:SoftModelShapeFunction}
 S_\mathrm{mod}(\mu) 
 & = & 
 \frac{128}{3} \frac{\mu^3}{\Lambda_m^4} \exp\left( - \frac{4\mu}{\Lambda_m} \right),
\eq
where $\Lambda_m$ is a smearing parameter, that is varied between 1 and 3 GeV.
The soft model shape function causes a smearing of the partonic thrust distribution that shifts the position of the peak to the positive direction. 
We denote by $\tau_\mathrm{peak}(Q_0)$ the peak position of the hadronic thrust distribution, where 
$\diff\sigma_{\mathrm{sing}}/\diff\tau$ has been substituted by $\diff\sigma_{\mathrm{shower}}/\diff\tau$.
We are interested in the peak position as a function of the shower cutoff $Q_0$, while all other parameters are kept fixed.
From soft-collinear effective theory (or the coherent branching formalism) one predicts \cite{Hoang:2018zrp}
\bq
\label{eq:TauPeakRelation}
  \tau_\mathrm{peak}(Q_0) 
  & = & 
  \tau_\mathrm{peak}(Q_0') - \frac{16}{Q} \int\limits_{Q_0'}^{Q_0} \diff R\, \frac{\alphaS(R)\CF}{4\pi}.
\eq
Here, $\tau_\mathrm{peak}(Q_0')$ is the value of the peak position at some reference scale $Q_0'$.

For physical meaningful results a change in the shower cutoff from $Q_0'$ to $Q_0$ should be accompanied by a modification of the 
soft model shape function $S_\mathrm{mod}$ according to
\bq
\label{cutoff_independence}
  \frac{\diff \sigma}{\diff \tau} (\tau,Q,Q_0) 
  & = &
  \int\limits_0^{\tau Q} \diff\mu\, \frac{\diff\sigma_{\mathrm{shower}}}{\diff\tau}\left(\tau-\frac{\mu}{Q},Q,Q_0'\right)
  S_\mathrm{mod}(\mu+\Delta_\mathrm{soft}(Q_0)-\Delta_\mathrm{soft}(Q_0')).
\eq
$\Delta_\mathrm{soft}$ is called the gap function \cite{Hoang:2007vb}.
The gap can be calculated perturbatively. 
The infrared insensitive difference between the gap function at the two scales yields to leading order
\begin{equation}
    \Delta_\mathrm{soft}(Q_0)-\Delta_\mathrm{soft}(Q_0') = 16 \int\limits_{Q_0'}^{Q_0} \diff R \frac{\alphaS(R)\CF}{4\pi}.
\end{equation}
The primary purpose of this paper is to verify that a dipole shower behaves as predicted by factorisation and soft-collinear effective theory.
Thus we keep the soft model shape function fixed and verify eq.~(\ref{eq:TauPeakRelation}).

On the other hand, if we correspondingly modify the soft model shape function according to eq.~(\ref{cutoff_independence}),
the peak position of the thrust distribution should be independent of the cutoff $Q_0$.
We also verify this relation.

\subsection{The massive case}

We now consider the massive case.
The leading order Born process is
$e^+ e^- \rightarrow Z/\gamma \rightarrow \bar{t} t$.
We denote by $m_{\mathrm{MC}}$ the Monte Carlo mass used in the hard matrix element and in the parton shower.
We assume the hierarchy
\bq
 Q_0 \; \ll \; m_{\mathrm{MC}} \; \ll \; Q.
\eq
As in ref.~\cite{Hoang:2018zrp} we modify the parameter $\Lambda_m$ entering the soft model shape function according to
\bq
\label{eq_Lambda_m_massive}
 \Lambda_m
 & \rightarrow &
 \Lambda_m
 + 4 \frac{m_{\mathrm{MC}} \Gamma_{\mathrm{MC}}}{Q}
\eq
to account for some additional smearing due to the top quark width $\Gamma_{\mathrm{MC}} = 1.5 \mathrm{GeV}$.
Again we study the peak position $\tau_\mathrm{peak}(Q_0)$ as a function 
of the shower cutoff $Q_0$, while keeping all other quantities fixed.
In the massive case, eq.~(\ref{eq:TauPeakRelation}) generalises to 
\bq
\label{eq:TauPeakRelationMassive}
  \tau_\mathrm{peak}(Q_0) 
  & = & 
  \tau_\mathrm{peak}(Q_0') - \frac{1}{Q} \left[ 16 -8 \pi \frac{m_{\mathrm{MC}}}{Q} \right] \CF \int\limits_{Q_0'}^{Q_0} \diff R\, \frac{\alphaS(R)}{4\pi}.
\eq
Note that in this case the shift in the peak position is less pronounced due to the alternating sign
in the square bracket.
For physical meaningful results a change in the shower cutoff should 
be accompanied by a modification of the 
soft model shape function as above and a redefinition of the Monte Carlo mass $m_{\mathrm{MC}}$:
\bq
 m_{\mathrm{MC}}\left(Q_0\right) - m_{\mathrm{MC}}\left(Q_0'\right)
 & = &
 - 2 \pi \CF \int\limits_{Q_0'}^{Q_0} \diff R\, \frac{\alphaS(R)}{4\pi}.
\eq
As before, we are in this paper primarily concerned to
verify that a dipole shower behaves as predicted by factorisation and soft-collinear effective theory.
Thus we keep the soft model shape function and  the Monte Carlo mass fixed 
and verify eq.~(\ref{eq:TauPeakRelationMassive}).

\section{Dipole shower results}
\label{sect:dipoleShower}

In this section we simulate events with the Nagy-Soper dipole shower algorithm \cite{Nagy:2005aa,Nagy:2006kb}. 
For the dipole shower algorithm we use the implementation of ref.~\cite{Dinsdale:2007mf}.
We first study the massless case, where we verify eq.~(\ref{eq:TauPeakRelation}) and eq.~(\ref{cutoff_independence}).
We then proceed to the massive case and verify eq.~(\ref{eq:TauPeakRelationMassive}).

As most showers, the implementation of ref.~\cite{Dinsdale:2007mf} is of leading-logarithmic accuracy (LL), meaning that the
shower correctly reproduces all leading-logarithmic terms (in the leading-colour approximation).
Leading-log parton showers include some, but not all next-to-leading logarithms (NLL).
The construction of NLL parton showers is a topic of current research \cite{Dasgupta:2018nvj,Forshaw:2020wrq,Dasgupta:2020fwr}.
Eqs.~(\ref{eq:TauPeakRelation}), (\ref{cutoff_independence}) and (\ref{eq:TauPeakRelationMassive}) give the leading terms
for the dependence on the cutoff parameter $Q_0$.
The dependence on the cutoff is approximately proportional to $(Q_0-Q_0')$ and shows up in LL and NLL terms.
Thus NLL accuracy is a sufficient, but not necessary condition for a parton shower to follow the predicted cutoff dependence from SCET.

\subsection{The massless case}

In the massless case we consider the hard process 
$e^+ e^- \rightarrow Z/\gamma \rightarrow \bar{q} q$ with massless quarks $q$ at the centre-of-mass energy $Q$, 
followed by the dipole parton shower with cutoff scale $Q_0$.
For the strong coupling we use the leading-order formula
\bq
 \alpha_s(\mu) & = & \frac{4\pi}{\beta_0 \ln \frac{\mu^2}{\Lambda^2}},
 \;\;\;
 \beta_0 = 11 - \frac{2}{3} N_f
\eq
with $\Lambda_5 = 88 \;\mbox{MeV}$ corresponding to $\alpha_s(m_Z) = 0.118$. 
For the centre-of-mass energy we consider $Q=m_Z$ and $Q=300 \, \mathrm{GeV}$.
For our analysis, we generate $10^7$ events for each cutoff scale $Q_0$ between $0.6 \, \mathrm{GeV}$ and $2.0 \, \mathrm{GeV}$ in steps of $0.2 \, \mathrm{GeV}$. 
The histograms are created with a bin size of $\Delta\tau = 10^{-3}$. 
Note that this procedure gives the partonic thrust distribution generated by the dipole parton shower algorithm. 
To make it comparable to the analytic results we convolve the partonic thrust distribution 
with the soft model shape function as in (\ref{eq:HadronicThrust}). 
For that purpose we convolve the partonic thrust distribution $\diff\sigma_{\mathrm{shower}}/\diff\tau$ 
with the $S_\mathrm{mod}$ of (\ref{eq:SoftModelShapeFunction}) using the discretised representation of the convolution integral.

\begin{figure}[t]
    \centering
    \includegraphics[width=0.55\textwidth]{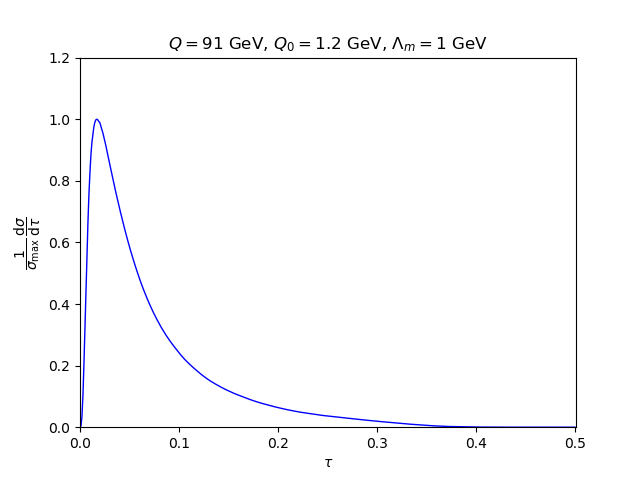}
    \caption{Thrust distribution for the centre-of-mass energy $Q=m_Z$, with the cutoff value $Q_0=1.2 \, \mathrm{GeV}$ 
and the smearing parameter of the soft model shape function $\Lambda_m=1 \, \mathrm{GeV}$. 
The distribution is normalized such that its peak value is one.}
    \label{fig:Thrust-91-12-1}
\end{figure}

As an example, we show the convolved partonic thrust distribution with the cutoff $Q_0=1.2 \, \mathrm{GeV}$ 
and the smearing parameter $\Lambda_m=1 \, \mathrm{GeV}$ for $Q=m_Z$ in fig.~\ref{fig:Thrust-91-12-1}. 
Note that we normalize all distributions in this paper such that the peak values are one.

Let us now address the question, how the position of the peak changes as a function of the cutoff $Q_0$, while keeping all other 
quantities fixed. In particular we keep the soft model shape function fixed. 

\begin{figure}[t!]
    \centering
    \begin{minipage}[t]{0.45\textwidth}
    \centering
    \includegraphics[width=1\textwidth]{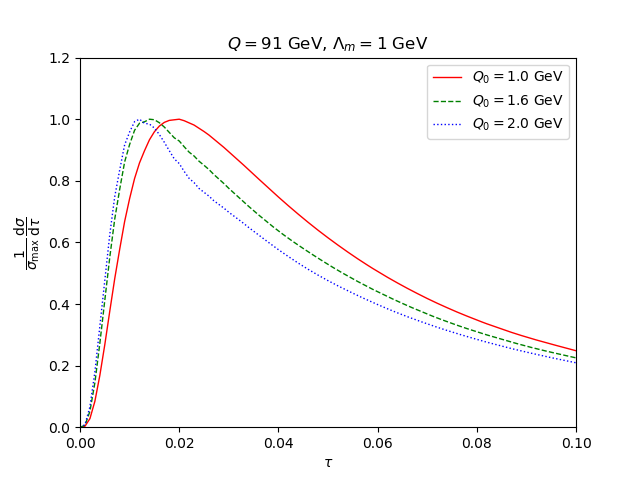}
    \end{minipage}
    \begin{minipage}[t]{0.45\textwidth}
    \centering
    \includegraphics[width=1\textwidth]{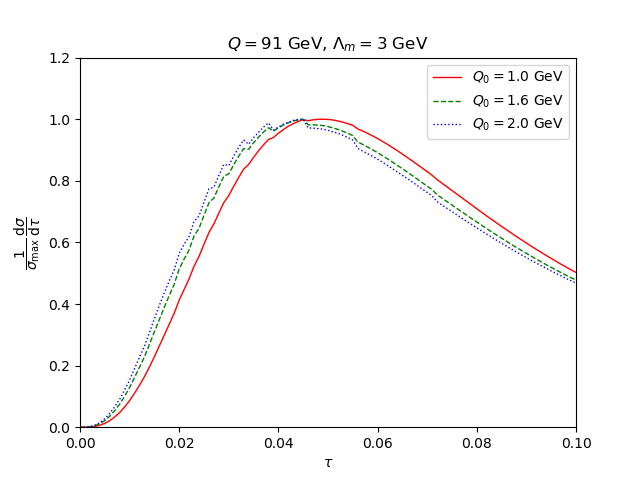}
    \end{minipage}
    
    \centering
    \begin{minipage}[t]{0.45\textwidth}
    \centering
    \includegraphics[width=1\textwidth]{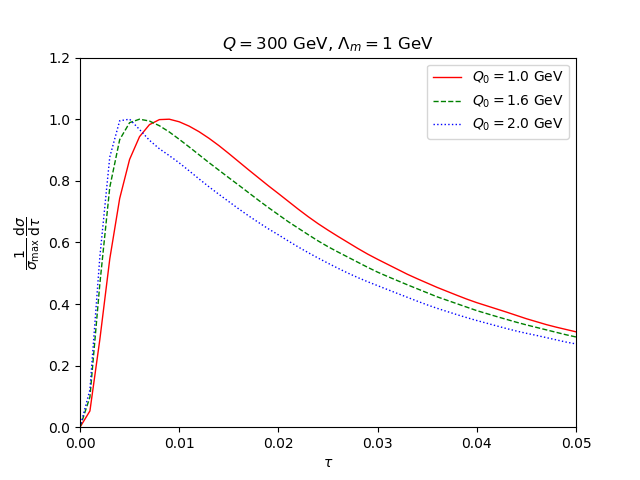}
    \end{minipage}
    \begin{minipage}[t]{0.45\textwidth}
    \centering
    \includegraphics[width=1\textwidth]{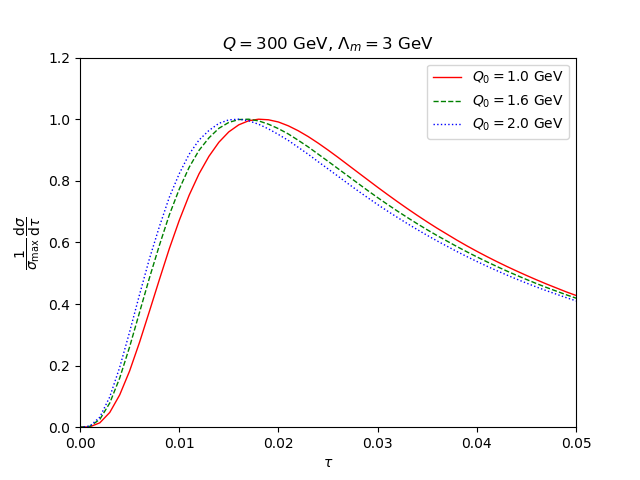}
    \end{minipage}
    \caption{Plots of the thrust distribution from the dipole shower 
for the smearing parameters $\Lambda_m=1 \, \mathrm{GeV}$ (left) and $\Lambda_m=3 \, \mathrm{GeV}$ (right) 
for the a centre-of-mass energies $Q=m_Z$ (upper) and $Q=300 \, \mathrm{GeV}$ (lower). 
Each figure shows the distribution for cutoff values of $Q_0=1 \, \mathrm{GeV}$ (red solid line), 
$Q_0=1.6 \, \mathrm{GeV}$ (green dashed line), and $Q_0=2 \, \mathrm{GeV}$ (blue dotted line), respectively.}
    \label{fig:Thrust_three_1and3}
\end{figure}

Fig.~\ref{fig:Thrust_three_1and3} shows the thrust distribution for the smearing parameters $\Lambda_m=1 \, \mathrm{GeV}$ (left column) 
and $\Lambda_m=3 \, \mathrm{GeV}$ (right column) with $Q=m_Z$ (upper panel) and $Q=300 \, \mathrm{GeV}$ (lower panel), 
each with three different values for the shower cut. 
The red solid line represents a cut of $Q_0=1 \, \mathrm{GeV}$. 
The green dashed line and the blue dotted line stand for the cutoff values $Q_0=1.6 \, \mathrm{GeV}$ and $Q_0=2 \, \mathrm{GeV}$, respectively. 
Already from the plots for the three different cutoff values we can recognize the tendency that is implied by the analytic prediction (\ref{eq:TauPeakRelation}): 
The value of the peak position decreases for larger cutoff values. 

\begin{figure}[t!]
\centering
\begin{minipage}[t]{0.45\textwidth}
 \includegraphics[width=1\textwidth]{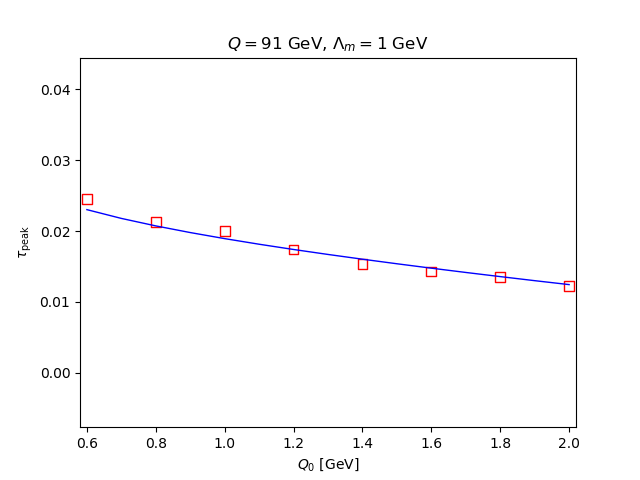}
\end{minipage}
\begin{minipage}[t]{0.45\textwidth}
 \includegraphics[width=1\textwidth]{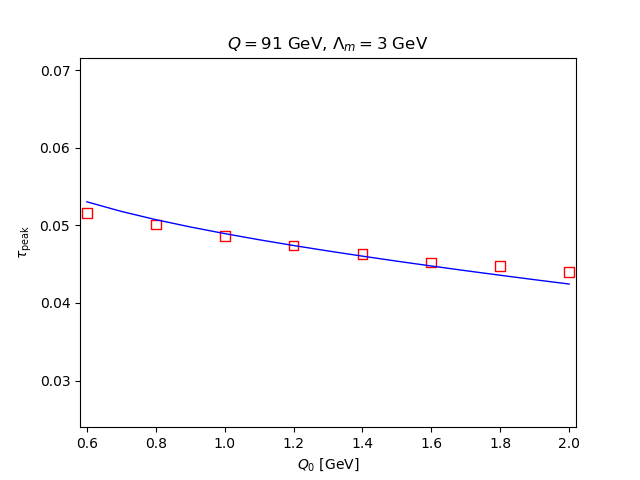}
\end{minipage}

\begin{minipage}[t]{0.45\textwidth}
 \includegraphics[width=1\textwidth]{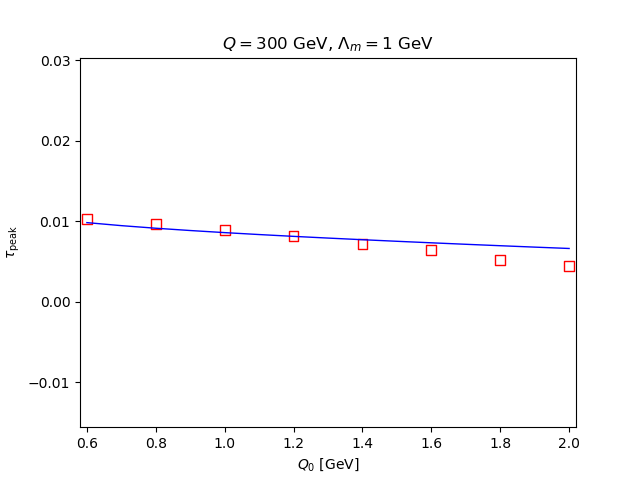}
\end{minipage}
\begin{minipage}[t]{0.45\textwidth}
 \includegraphics[width=1\textwidth]{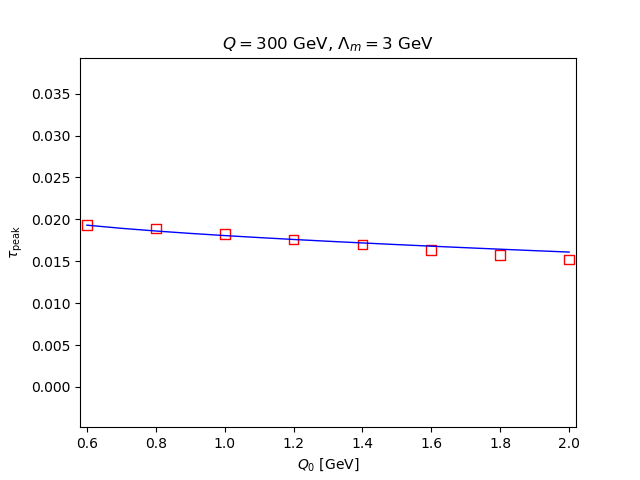}
\end{minipage}
\caption{Position of the peak of the thrust distribution as a function of the shower cutoff $Q_0$. 
Displayed are the plots for $Q=m_Z$ with $\Lambda_m=1 \, \mathrm{GeV}$ (upper left) and $\Lambda_m=3 \, \mathrm{GeV}$ (upper right) 
as well as $Q=300 \, \mathrm{GeV}$ with $\Lambda_m=1 \, \mathrm{GeV}$ (lower left) and $\Lambda_m=3 \, \mathrm{GeV}$ (lower right). 
The blue solid line depicts the result of the analytical computation while the red squares represent the data points from the parton shower simulation.}
\label{fig:CutoffDependence}
\end{figure}

To obtain a more quantitative analysis of the peak position's dependence on the cutoff, 
we fit a quadratic function to the thrust distribution in the peak region and extract the maximum. 
This procedure ensures that the statistical uncertainties in the determination of the peak position are so small 
that we desist from specifying any systematic or statistical errors in our results. 
The results from this analysis are shown in fig.~\ref{fig:CutoffDependence}. 
The plot of the peak position against the cutoff $Q_0$ from the analytic formula 
is obtained by solving the integral in (\ref{eq:TauPeakRelation}) numerically.
As reference value we take the peak position from the parton shower simulation at $Q_0'=1.2 \, \mathrm{GeV}$.

Fig.~\ref{fig:CutoffDependence} illustrates the position of the peak $\tau_\mathrm{peak}$ as a function of the cutoff $Q_0$ 
for four different combinations of $Q$ and $\Lambda_m$. 
The two upper figures correspond to $Q=m_Z$ with $\Lambda_m=1 \, \mathrm{GeV}$ (left) and $\Lambda_m=3 \, \mathrm{GeV}$ (right), 
while the two lower figures depict the $Q_0$ dependence for $Q=300 \, \mathrm{GeV}$ with $\Lambda_m=1 \, \mathrm{GeV}$ (left) 
and $\Lambda_m=3 \, \mathrm{GeV}$ (right). 
The blue solid line shows the analytic relation (\ref{eq:TauPeakRelation}). 
The centres of the red squares are the data points obtained from the parton shower simulations.

From fig.~\ref{fig:CutoffDependence} we deduce a good agreement between the analytical prediction 
and the parton shower simulations from the dipole shower formalism. Hence, our results coincide with the findings of \cite{Hoang:2018zrp}.

In the plots above we verified that the shower behaves as expected. For physical observables a change in the shower cutoff should be
accompanied by a modification of the soft model shape function as in eq.~(\ref{cutoff_independence}).

\begin{figure}[t]
    \centering
    \includegraphics[width=0.55\textwidth]{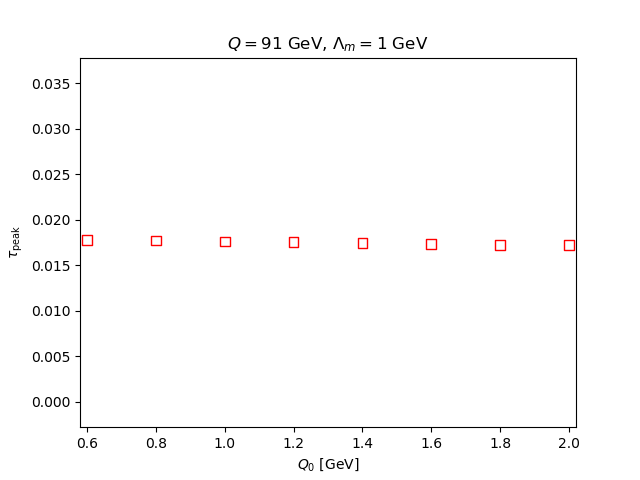}
    \caption{Position of the peak of the thrust distribution as a function of the shower cutoff $Q_0$ when the soft model shape function is modified
accordingly. 
Displayed is a plot for $Q=m_Z$ with $\Lambda_m=1 \, \mathrm{GeV}$.}
    \label{fig:tau_peak_vs_Q0_91_1}
\end{figure}

In fig.~\ref{fig:tau_peak_vs_Q0_91_1} we show the position of the peak of the thrust distribution as a function of the shower cutoff $Q_0$,
where we modify the soft model shape function according to eq.~(\ref{cutoff_independence}).
We expect that the leading terms cancel in the combination and indeed fig.~~\ref{fig:tau_peak_vs_Q0_91_1} shows that the result
is independent of the cutoff $Q_0$.
In fig.~\ref{fig:tau_peak_vs_Q0_91_1} the results for $Q=m_Z$ and $\Lambda_m=1 \, \mathrm{GeV}$ are displayed.

\subsection{The massive case}

We now turn to the massive case. We consider 
the hard process 
$e^+ e^- \rightarrow Z/\gamma \rightarrow \bar{t} t$ at the centre-of-mass energy $Q$, 
followed by the dipole parton shower with cutoff scale $Q_0$.
The analytic results are derived under the assumption of the hierarchy
\bq
 Q_0 \; \ll \; m_{\mathrm{MC}} \; \ll \; Q.
\eq
Thus we consider $m_{\mathrm{MC}} = 173 \, \mathrm{GeV}$ and $Q = 1000 \, \mathrm{GeV}$.
The original implementation of \cite{Dinsdale:2007mf} uses in the massive case the evolution variable
\bq
 t & = & \ln \frac{-p_\perp^2+(1-z)^2 m_i^2 + z^2 m_j^2}{Q^2},
\eq
as suggested by \cite{Dokshitzer:1995qm,Gieseke:2003rz}.
$m_i$ and $m_j$ are the masses of the emitter and the emitted particle after the splitting.
For this publication we changed the evolution variable to
\bq
 t & = & \ln \frac{-p_\perp^2}{Q^2}.
\eq
Thus all numerical results in this paper are obtained by using the transverse momentum as the shower evolution variable.
The plots in the massive case are based on $10^5$ events for each cutoff scale $Q_0$.
The cutoff scale is again varied between $0.6 \, \mathrm{GeV}$ and $2.0 \, \mathrm{GeV}$ in steps of $0.2 \, \mathrm{GeV}$. 
We study again the peak position as a function of the cutoff $Q_0$, while keeping all other 
quantities fixed.
In particular we keep the soft model shape function and the Monte Carlo mass fixed. 
From eq.~(\ref{eq:TauPeakRelationMassive}) we expect the shift in the peak position to be less pronounced,
mainly due to the larger centre-of-mass energy, which enters as a $1/Q$-prefactor but also due to the alternating sign
in the factor $16 - 8\pi m_{\mathrm{MC}}/Q$.

In the soft model shape function we use
\bq
 \Lambda_m & = &
  \Lambda_{m,0} + 4 \frac{m_{\mathrm{MC}} \Gamma_{\mathrm{MC}}}{Q},
\eq
as indicated by eq.~(\ref{eq_Lambda_m_massive}).

\begin{figure}[t!]
\centering
\begin{minipage}[t]{0.45\textwidth}
 \includegraphics[width=1\textwidth]{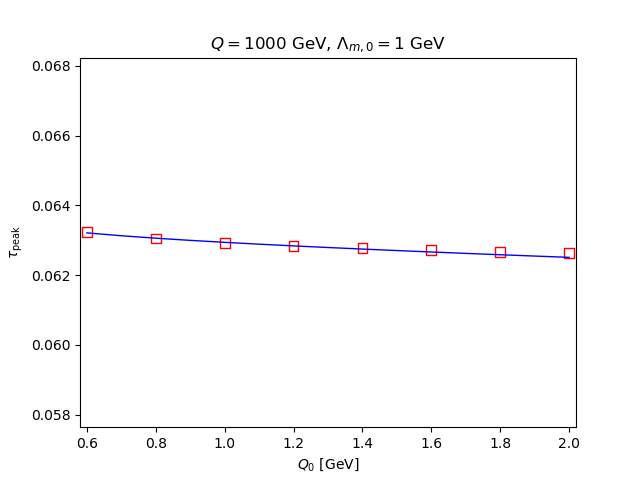}
\end{minipage}
\begin{minipage}[t]{0.45\textwidth}
 \includegraphics[width=1\textwidth]{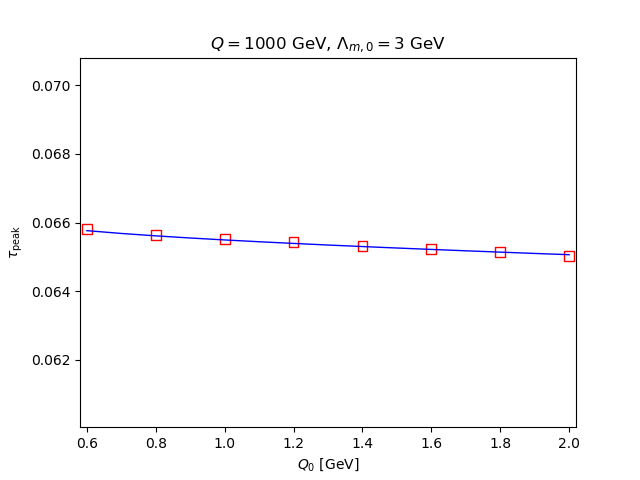}
\end{minipage}
\caption{Position of the peak of the thrust distribution as a function of the shower cutoff $Q_0$ for $e^+ e^- \rightarrow \bar{t} t$.
The centre-of-mass energy is $Q = 1000 \, \mathrm{GeV}$.
Displayed are the plots for $\Lambda_{m,0}=1 \, \mathrm{GeV}$ (left) and $\Lambda_{m,0}=3 \, \mathrm{GeV}$ (right).
The blue solid line depicts the result of the analytical computation while the red squares represent the data points from the parton shower simulation.}
\label{fig:Massive}
\end{figure}

Fig.~\ref{fig:Massive} shows our results.
The centre-of-mass energy is $Q = 1000 \, \mathrm{GeV}$.
For the top quark we take $m_{\mathrm{MC}} = 173 \, \mathrm{GeV}$ and $\Gamma_{\mathrm{MC}} = 1.5 \, \mathrm{GeV}$.
The left figure corresponds to $\Lambda_{m,0}=1 \, \mathrm{GeV}$ 
while the right figure depicts $\Lambda_{m,0}=3 \, \mathrm{GeV}$.
The blue solid line shows the analytic relation (\ref{eq:TauPeakRelationMassive}). 
The centres of the red squares are the data points obtained from the parton shower simulations.
We observe a good agreement between the analytical prediction and the numerical shower result.

\section{Conclusions}
\label{sect:conclusions}

In this paper we analysed the cutoff dependence of the thrust peak position using an implementation of the Nagy-Soper dipole parton shower. 
We did this for the massless case and the massive case.
We showed that the results of the numerical shower simulation agree with the analytic relations found in \cite{Hoang:2018zrp}. 
The analytic relations have been derived within soft-collinear effective theory (or alternatively the coherent branching formalism).
This is an important verification. It shows that the results of \cite{Hoang:2018zrp} are not restricted
to an angular ordered shower, but apply to a dipole shower with transverse momentum as evolution variable as well.
We expect the results to hold for any parton shower whose shower evolution variable reduces in the singular limit to $p_\perp$.
In particular we expect our results to apply to the default shower of the Sherpa event generator \cite{Schumann:2007mg},
as this shower is an independent implementation of the same shower algorithm we are using. 

More generally, given an observable, a parton shower and an effective theory describing adequately the physics 
at the scales of the parton shower we expect that an observable like the shift of the peak position 
can systematically be calculated
in the effective theory for the given parton shower algorithm.

Turning to phenomenology, the results of this paper improve our understanding of the Monte Carlo mass, which in turn is important for the determination of the top quark mass.

\subsection*{Acknowledgements}

We thank Simon Pl\"atzer for useful comments and discussions.

This work has been supported through the BMBF project
``Precision calculations for collider and Higgs physics at the LHC'' (Project ID 05H18UMCA1).


{\footnotesize
\bibliography{/home/stefanw/notes/biblio}
\bibliographystyle{/home/stefanw/latex-style/h-physrev5}
}

\end{document}